\documentclass{article}
\usepackage[utf8]{inputenc}
\usepackage{cite}
\usepackage{amsmath,amssymb,amsfonts}
\usepackage{algorithmic}
\usepackage{graphicx}
\usepackage{textcomp}
\usepackage{hyperref}
\usepackage{multirow}
\usepackage{authblk}
\def\BibTeX{{\rm B\kern-.05em{\sc i\kern-.025em b}\kern-.08em
    T\kern-.1667em\lower.7ex\hbox{E}\kern-.125emX}}
\newtheorem{example}{Example}
\begin{document}
%\history{Date of publication xxxx 00, 0000, date of current version xxxx 00, 0000.}
%\doi{10.1109/ACCESS.2017.DOI}

\title{Generating Biomedical Question Answering Corpora from Q\&A Forums}

\author[1]{Andre Lamurias}
\author[1]{Diana Sousa}
\author[1]{Francisco M Couto}

\affil[1]{LASIGE, Departamento de Informática, Faculdade de Ciências, Universidade de Lisboa, 1749-016, Lisbon, Portugal}

%\begin{keywords}
%question answering, document retrieval, text mining, datasets, deep learning
%\end{keywords}

%\titlepgskip=-15pt

\maketitle

\begin{abstract}
Question Answering (QA) is a natural language processing task that aims at obtaining relevant answers to user questions.
While some progress has been made in this area, biomedical questions are still a challenge to most QA approaches, due to the complexity of the domain and limited availability of training sets.
We present a method to automatically extract question-article pairs from Q\&A web forums, which can be used for document retrieval, a crucial step of most QA systems.
The proposed framework extracts from selected forums the questions and the respective  answers that contain citations.
This way, QA systems based on document retrieval can be developed and evaluated using the question-article pairs annotated by users of these forums.
We generated the BiQA corpus by applying our framework to three forums, obtaining 7,453 questions and 14,239 question-article pairs.
We evaluated how the number of articles associated with each question and the number of votes on each answer affects the performance of baseline document retrieval approaches.
Also, we demonstrated that the articles given as answers are significantly similar to the questions and trained a state-of-the-art deep learning model that obtained similar performance to using a dataset manually annotated by experts.
The proposed framework can be used to update the BiQA corpus from the same forums as new posts are made, and from other forums that support their answers with documents.
The BiQA corpus and the framework used to generate it are available at \url{https://github.com/lasigeBioTM/BiQA}.
\end{abstract}

\section{Introduction}
\label{sec:introduction}

Question Answering (QA) consists of the automatic retrieval of information that can directly answer user questions.
This task is relevant for the biomedical domain, due to the large quantity and variety of information that is necessary to integrate to fully understand biomedical problems.
QA systems can assist the study of biological systems by automatically retrieving information relevant to the queries made by a researcher.
This is possible because of all the biomedical information that is available in text format on publicly available digital libraries, in the form of scientific literature.
However, these systems require advanced techniques when dealing with large document repositories.

In recent years, many developments have been made in QA systems using deep learning techniques~\cite{Pang2017,guo2016deep}.
QA systems are developed using gold standards, where automatic answers are compared to the manually annotated answers by domain experts.
One of the most relevant gold standards for biomedical QA is BioASQ~\cite{tsatsaronis2015overview}, which is also a community challenge where every year, a new dataset is released to train and evaluate QA systems.
One limitation of this type of evaluation is that the questions are not obtained from real users.
The BioASQ datasets are developed by experts, who create the questions and the answers.
Therefore it lacks the variability that is inherent to real-world scenarios.
furthermore, it requires the involvement of domain experts, which may not be possible in some cases.

Question-and-Answers (Q\&A) forums, such as Stack Exchange and Quora, are websites where users can post questions and answer the questions of other users.
In recent years, search engines have improved their ability to answer natural language queries, instead of relying just on keywords as input.
It is more intuitive for users to retrieve information using natural language than through keyword-based search.
QA systems could, therefore, benefit from Q\&A forums where users collaborate in posting and answering questions.

Although most of these forums have no selection process to restrict the participants, the users can vote on each answer, and, in some cases, the user who made the question can give their approval to one of the answers.
One significant example of Q\&A forums is Stack Exchange, which is a network of websites where users can ask questions about specific subjects, such as computer science, mathematics, art, and various languages.
Reddit is another forum where users can post questions and get answers from communities focused on specific subjects, such as nutrition.
Our work focused on Stack Exchange and Reddit since both these forums contain questions relevant to the biomedical domain, and provide an API to retrieve their contents.

For biomedical questions, it is essential to provide proper references that corroborate the given answers. 
Otherwise, the correctness of the answer may be questioned.
Usually, Q\&A forums that are related to biomedical subjects, including the ones used in this work, encourage users to support their posts with scientific articles.
A significant advantage of open-access digital libraries such as PubMed is that anyone can analyze its documents to better understand biomedical problems.
For this reason, improving biomedical document retrieval is of interest to the general community.

With this article, we present a framework to generate a gold standard for document retrieval QA, with questions and relevant articles retrieved automatically from Q\&A forums.
We demonstrated our approach on three forums that are relevant to biomedical sciences: Biology from Stack Exchange, Medical Sciences also from Stack Exchange, and Nutrition from Reddit.

We also present the BiQA corpus of question-article pairs, which can be used to develop and evaluate document retrieval and QA systems, as we demonstrate in this article.
Unlike existing corpora, our corpus is based on real user questions and can evolve over time.
The proposed framework could also be applied to other relevant forums, in any language, and update the dataset with new questions.
To the best of our knowledge, this is the first work to automatically obtain document-based QA training data from web forums.

\section{Related Work}

\subsection{User-based QA}

Recently, there has been an increased focus on QA using real user questions.
For example, the Natural Questions dataset released by Google uses real queries made by users of the Google search engine \cite{kwiatkowski2019natural}.
While the questions of other QA datasets are formulated by annotators whose function is to read a paragraph and write relevant questions \cite{rajpurkar-etal-2018-know}, this dataset contains real queries from users seeking information.
The authors then annotated each question with answers taken from the English Wikipedia.
The objective of this dataset is to develop systems that can extract the correct answer to each question from a specific Wikipedia page.

Another example is the TREC LiveQA track~\cite{AgichteinCPPH15}, which promotes the development of new approaches to answer questions retrieved from various sources, such as Yahoo Answers and the U.S. National Library of Medicine.
The results of the teams that have participated in this track highlight the low scores obtained when compared with human performance, particularly for consumer health questions~\cite{AbachaAPD17}.

\subsection{Datasets based on web forums}
Other authors have also explored Q\&A forums to generate datasets.
\cite{Cong2008} presented an algorithm to extract Q\&A pairs from forum threads, where it may not be clear which posts are questions and which are answers.
\cite{Shah:2010:EPA:1835449.1835518} generated a dataset based on Yahoo Answers and used crowdsourcing to rank the answers given to the same question.
However, since Stack Exchange and Reddit users can vote on answers for any question (not just the ones they made), our approach does not require that extra annotation step.
\cite{Dalip2013} explored Stack Overflow to develop a learning-to-rank method to evaluate the quality of the answers, based on the scores given by the users.
The authors extracted several textual and user-based features to classify each answer, including the number of links to external sources, which they found to be a good predictor of the quality of the answer.
Nevertheless, to the best of our knowledge, we were the first to extract references to articles from the answers to automatically generate a QA dataset.
\cite{gupta2019amazonqa} demonstrated an approach to review-based QA using questions posted on Amazon product reviews.
In this study, they highlight the difficulty of answering user questions in a real-world scenario due to many questions not being answerable.
However, they modeled this task as a closed-world QA task, where it is known which snippets contain the answers.
In our case, we consider an open-world scenario where any document from the biomedical literature can contain the answer.

\subsection{Biomedical QA}

The development of systems for biomedical QA requires gold standard datasets to both train and evaluate those systems.
The biomedical domain poses a challenge to this type of system since the questions may be more challenging to answer, and the datasets available are more limited.
BioASQ is one of the most significant community efforts in advancing biomedical QA systems.
It is a series of evaluations where the participants compete on various QA tasks, using a gold standard provided by the organization.
Each edition of BioASQ has had at least these two subtasks: one which consists of indexing PubMed abstracts with MeSH terms, while the second consists of answering biomedical questions.
Each year the organizers provide new annotated questions that are curated by biomedical experts.
In 2019 (BioASQ 7), task B phase A consisted of retrieving relevant documents, concepts, and snippets from designated article repositories to answer biomedical questions.
The best Mean Average Precision (MAP) scores ranged between 0.1218 and 0.2898 on five test batches.
This dataset is similar to ours because PubMed abstracts are used to answer natural language questions.
However, while the BioASQ questions are created and answered by experts, our dataset is made of questions and answers created by users of online forums.
The consequence of this difference is that our questions are taken directly from real users, and the answers are given by non-experts, which are easier to obtain.

There have been other efforts in developing datasets for biomedical QA tasks.
MediQA~\cite{mediqa2019-overview} is another relevant challenge, which focuses on clinical data.
This dataset consists of question and candidate answers, which the participants had to classify and rank according to the relevance to the question.
While this dataset was developed manually to include both correct and incorrect answers, the organizers also provided other datasets derived from online resources which contain only the correct answers~\cite{abacha2019question}.
The dataset released for this task contained 383 questions, while the answers were retrieved using a QA system and ranked by experts.
emrQA is another dataset that also focused on the clinical domain~\cite{pampari-2018-emrqa}.
It consists of 6,686 questions and explores the i2b2 dataset to generate QA pairs.
Recently, \cite{jin2019pubmedqa} released the PubMedQA dataset, where they derive questions based on article titles and answer them with their respective abstracts.

\section{Corpus generation}

\begin{figure*}
\centering
\includegraphics[width=0.5\linewidth]{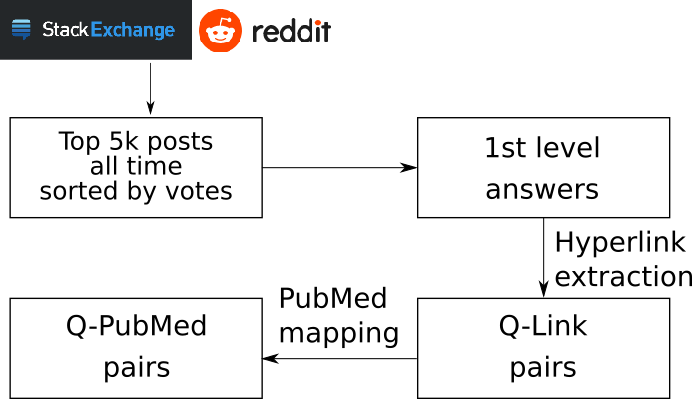}
\caption{Framework used to generate the BiQA corpus.} \label{fig:pipeline}
\end{figure*}

Fig. \ref{fig:pipeline} summarizes our framework that we used to generate the BiQA corpus from Q\&A forums.
For each post of a forum, we retrieved its title, body text, and the number of votes.
If the forum is not specifically for Q\&A, such as Reddit, we selected only posts with question marks in the title or body.
For each post, we retrieved all of its first-level replies, i.e., we ignored comments made on the question and answers because these usually are not direct answers to the question. 
We also did not take into account user data, although other authors have pointed out that that information could be relevant to rank answers \cite{Dalip2013}.
However, we saved the original IDs of every post and reply so that additional information could be retrieved later.

Afterward, we parsed each answer to obtain hyperlinks mentioned throughout the answer text.
Whenever possible, we linked various types of URLs to PubMed: from PMC, doi.org, ScienceDirect, and ResearchGate.
More types of URLs could be handled in the future using more complex mapping rules or using machine learning.
Other users provided the full formatted citation without a hyperlink, which would require more complex methods to link to PubMed automatically.

Example \ref{ex1} shows a question from our corpus, and a document used to answer it.
The answer has some text written by the user as well as the passage from the article that contained the answer, which we omitted.
This is a simple example where the user admits that they are not experts and simply used academic search engines.
Indeed, if we search the question title text on PubMed, the article linked can be found in the results list.
Example \ref{ex2} provides another example where domain knowledge is necessary to retrieve the articles to answer the question.
In this case, the question text should be processed to remove stop words, and it is necessary to have some understanding of the relationship between cholesterol levels and consumption of eggs.
\\

\begin{example}
Question posted on Biology.SE where a user based their answer on an article retrieved from PubMed.
\begin{itemize}

\item \textbf{Question link:} \url{https://biology.stackexchange.com/questions/84829} \\

\item \textbf{Question title:} Number of dopaminergic neurons in VTA \\

\item \textbf{Answer text:} Approximately, 250000-440000 neurons in Humans (Rice et al., 2016). \\

\item \textbf{Answer PMID:} 25269834 (Mapping dopaminergic deficiencies in the substantia nigra/ventral tegmental area in schizophrenia.) \\
 \label{ex1}
\end{itemize}
\end{example}

\begin{example}
Question posted on MS.SE where a user based their answer on several biomedical articles (we show only one of those articles).
\begin{itemize}

\item \textbf{Question link:} \url{https://medicalsciences.stackexchange.com/questions/529} \\

\item \textbf{Question title:} How many eggs can one eat per day? \\

\item \textbf{Answer text:} (...) The two studies cited indicate that consumption of eggs does not raise cholesterol levels significantly for about 3/4ths of the population. (...) \\

\item \textbf{Answer PMID:} 16340654 (Dietary cholesterol provided by eggs and plasma lipoproteins in healthy populations.) \\
\label{ex2}
 
\end{itemize}
\end{example}

We did not filter the answers by their score; however, in our analysis, we tested various thresholds for the minimum number of votes and its effect on document retrieval engines (Section \ref{sec:data}).
When an answer has more votes, it means that more users agreed that that answer is correct or at least relevant.
While we could have applied stricter criteria for answer selection, we observed some answers supported by citations but without additional votes (every answer starts with one vote).
This is possibly due to the limited size of the forums' user base.
However, there were cases of negative scores, which can indicate that the answer is incorrect or non-related to the question, according to the users that voted.

\section{Corpus evaluation}

We evaluated the BiQA corpus using two strategies.
First, we compared the results obtained with three document search methods on the questions of the dataset.
This way, we could compare the references obtained through user answers with the ones obtained directly using search engines.
The scores obtained indicate if search engine techniques can obtain the same answers that were provided by the users.
Secondly, we evaluated the dataset by applying it to a biomedical question answering task.
The objective was to determine if, with our dataset, we can obtain similar performance to a dataset developed by domain experts.

\subsection{IR-based validation}
\label{sec:comparison}

In order to understand the potential of our dataset for document retrieval, we evaluated three robust baseline methods. 
We considered the articles obtained from the forum answers as the list of relevant documents of each question.
The three methods were: i) the NCBI Eutils API (\url{https://eutils.ncbi.nlm.nih.gov/entrez/eutils}) which provides access to the PubMed search engine, ii) query likelihood scorer with Dirichlet term smoothing calculated by Galago~\cite{croft2010search} on our local version of PubMed, and iii) BM25 scorer~\cite{robertson1999okapi}, also calculated by Galago.
The PubMed search engine also uses BM25 as a first stage, with additional improvements~\cite{fiorini2018best}. 

\begin{figure*}
\centering
\includegraphics[width=0.5\linewidth]{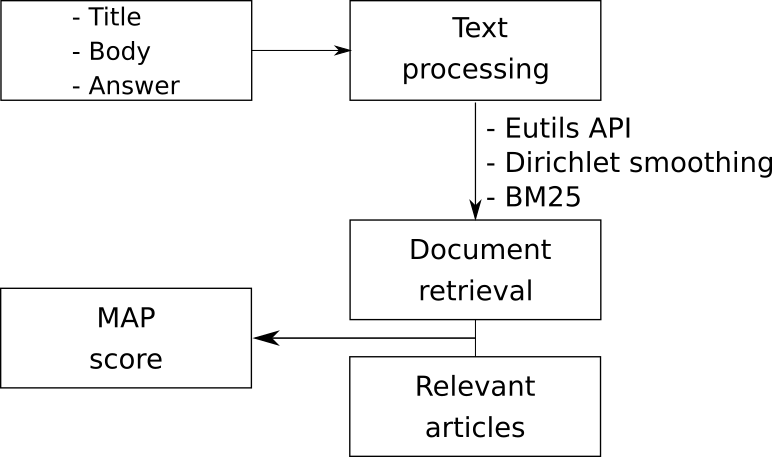}
\caption{Baseline evaluation of the BiQA corpus.} \label{fig:evaluation}
\end{figure*}

We tested querying each method with the title and body text of the questions, as well as with the answer text since the articles were usually more related to the answer text than to the question text (Fig.~\ref{fig:evaluation}).
Although in a QA task the answer text would not be available, we still wanted to study, if given the answer text, we could retrieve the relevant articles.

Our framework first tokenized the text using Spacy~(\url{https://spacy.io/}) and then removed HTML tags, stopwords, punctuation, and spaces.
We used the ``OR'' operator with the PubMed search since otherwise, the articles would have to contain the same words as the query, and we would get very few results.
We retrieved a maximum of 100 articles per question, ordered by the relevance score provided by the PubMed API, and the Dirichlet term smoothing score and BM25 score calculated by Galago.
We retrieved 100 articles because this was the same number of articles retrieved by \cite{nentidis2017results} (see Section \ref{sec:bioasq}).
Then we compared with the articles given by the users that answered each question, which we considered as correct if they had more than a given number of votes.
If the correct articles appear at the top of the search results, it means that they could have been obtained through a document search engine.
However, if the correct articles appear at the bottom or do not appear at all, it means that the document retrieval process fails for this type of question.

\subsection{BioASQ6b validation}
\label{sec:bioasq}
Document retrieval is a significant step of a QA pipeline since the answers are usually derived from the retrieved documents.
For this reason, we used our dataset with a biomedical QA system that performed document retrieval tasks to evaluate the effect on its performance.
We chose the AUEB system developed for BioASQ6~\cite{nentidis2017results} since it obtained the highest MAP scores on that competition for the document retrieval tasks.
The BioASQ6 dataset, where the AUEB system was evaluated, consists of a training set of 2,151 questions, a development set of 100 questions, and five testing batches of 100 questions each.
On this dataset, the results of each testing batch are reported separately. 

The AUEB team explored various deep learning models for BioASQ6b.
We selected their extension of the Position-Aware Convolutional Recurrent Relevance (PACRR)  model~\cite{Hui2017PACRRAP}, dubbed TERM-PACRR, as it obtained better results in most test batches for document retrieval.
This model computes a query-document similarity matrix and uses a convolution neural network and multi-layer perceptron to compute the relevance score of each query term.

We added the BiQA corpus to the BioASQ corpus to determine if the additional questions would bring improvements to the performance of the AUEB system.
To accomplish this, we first executed the AUEB system with the original BioASQ6 data to reproduce the results.
Then we reran it using the full BiQA dataset, in addition to the BioASQ data.
The AUEB system first retrieves 100 documents for each question using Galago and uses the BM25 score to then rerank these documents.
Therefore we did the same for the questions of the BiQA corpus, using our local version of PubMed indexed by Galago.
At each epoch, the model was evaluated on the development set, and, at the end of each training run, the model that obtained the best performance on the development set was selected.
We followed the procedure described by the original AUEB submission, which consisted in training ten versions of each model using different random seeds and combining the results of the ten models on each test batch.
The purpose of this procedure is to reduce the variability of the results of different training runs.

Finally, we also experimented only using the BiQA corpus to train a model with the AUEB system.
We followed the same procedure to evaluate this version of the model.

\subsection{Evaluation measures}

We calculated the Mean Average Precision (MAP) on each subset of the BiQA corpus and each validation method, which is also the metric used on BioASQ Task B Phase A.
This way, we can indirectly compare the scores obtained on our dataset with those obtained on the BioASQ datasets.
To calculate the MAP score, first, we calculate the average precision obtained with each question, and then the macro-average over all questions.
The average precision is given by:
\begin{equation}
    AP = \frac{\sum_{r=1}^{|L|} P(r) \cdot rel(r)}{|L_{R}|}
\end{equation}
where $|L|$ is the number of retrieved articles and $|L_{R}|$ is the number of relevant articles obtained from the answers, $P(r)$ is the precision obtained with the first $r$ retrieved articles, and $rel(r)$ is 1 if the $r$th retrieved article is relevant, and 0 otherwise. 
Considering Example~\ref{ex1}, since it only has one relevant article, its $AP$ would be equal to $\frac{1}{n}$ where $n$ is the position of the relevant article in the retrieved list.
For the BioASQ6b validation, we used a variation of this measure where $L_{R}$ is always equal to 10, which is how it is implemented on the evaluation tool.
Then we calculate MAP as:
\begin{equation}
    MAP = \frac{\sum_{i=1}^{|Q|}AP(q_{i})}{|Q|}
\end{equation}
where $q_{i} \in Q$ and $Q$ is the list of all questions.

\section{BiQA corpus analysis}

\subsection{Data sources}
We focused on three online forums to extract QA data: Biology and Medical Sciences from Stack Exchange (we refer to them as Biology.SE and MS.SE) and Nutrition from Reddit (r/nutrition).
We assumed that the users of these communities would give more value to answers supported by scientific articles, instead of accepting answers that could be based on personal experiences, for example.
On Reddit, only an account is necessary to create and answer posts, as well as to vote on posts and replies.
However, r/nutrition does not allow posts from new accounts or accounts with negative karma (balance of positive and negative votes).
On Stack Exchange, it is possible to answer as a guest without creating an account, but it has a reputation system so that only established users can vote on other answers.
Both forums have volunteer moderators who define community guidelines and can delete posts.
Table~\ref{table:communities} provides some statistics on these forums, obtained at the time of writing.

\begin{table}[ht!]
\centering
\caption{Statistics of the Q\&A forums used to generate the BiQA corpus. Reddit does not provide the total number of posts and the number of daily visitors.}
\label{table:communities}
\begin{tabular}{llllll}
\hline 
Forum stats & \#Posts & Users & Visits/day & Posts/day & Age  \\  \hline 
Biology.SE  & 24k  & 44k  & 10k  & 11 & 8y6m  \\ 
MS.SE       & 6.5k & 15k  & 2.8k & 3.9 & 5y3m  \\ 
r/nutrition & -    & 887k & -    & 20  & 11y  \\ \hline
\end{tabular}
\end{table}

\subsection{Corpus statistics}
\label{sec:data}

The BiQA corpus consists of a collection of three datasets, obtained from each of the forums previously described.
We show the number of questions and question-article pairs of each dataset in Table \ref{table:corpus}.
We considered only questions with answers that contained citations that we could automatically extract and map to PMIDs.
As we did not normalize the number of votes, we can see a relation between the average number of votes and size of the community; r/nutrition has a larger community than Biology.SE and MS.SE; hence the average number of votes is higher.

\begin{table*}[ht]
\center
\caption{Statistics of the three datasets that compose the BiQA corpus, along with the total number of questions and Q-PMID pairs and macro-averages.}\label{table:corpus}
\begin{tabular}{llllll}
\hline
Forum & Query date         & Qs w/PMID & Q-article pairs & Avg \# votes & Avg \# PMID       \\ \hline 
Biology.SE  & 04/05/2020 & 3961               & 6925         & 4.96            & 1.62                 \\ %\hline
MS.SE       & 04/05/2020 & 1383               & 3053         & 3.91            & 2.06                  \\ %\hline
r/nutrition & 04/05/2020 & 2109               & 4261         & 6.13            & 1.79                  \\ %\hline
Total       & -          & 7396               & 14133        & 5.01            & 1.82                 \\ \hline
\end{tabular}
\end{table*}

\begin{figure*}
\centering
\includegraphics[width=0.8\linewidth]{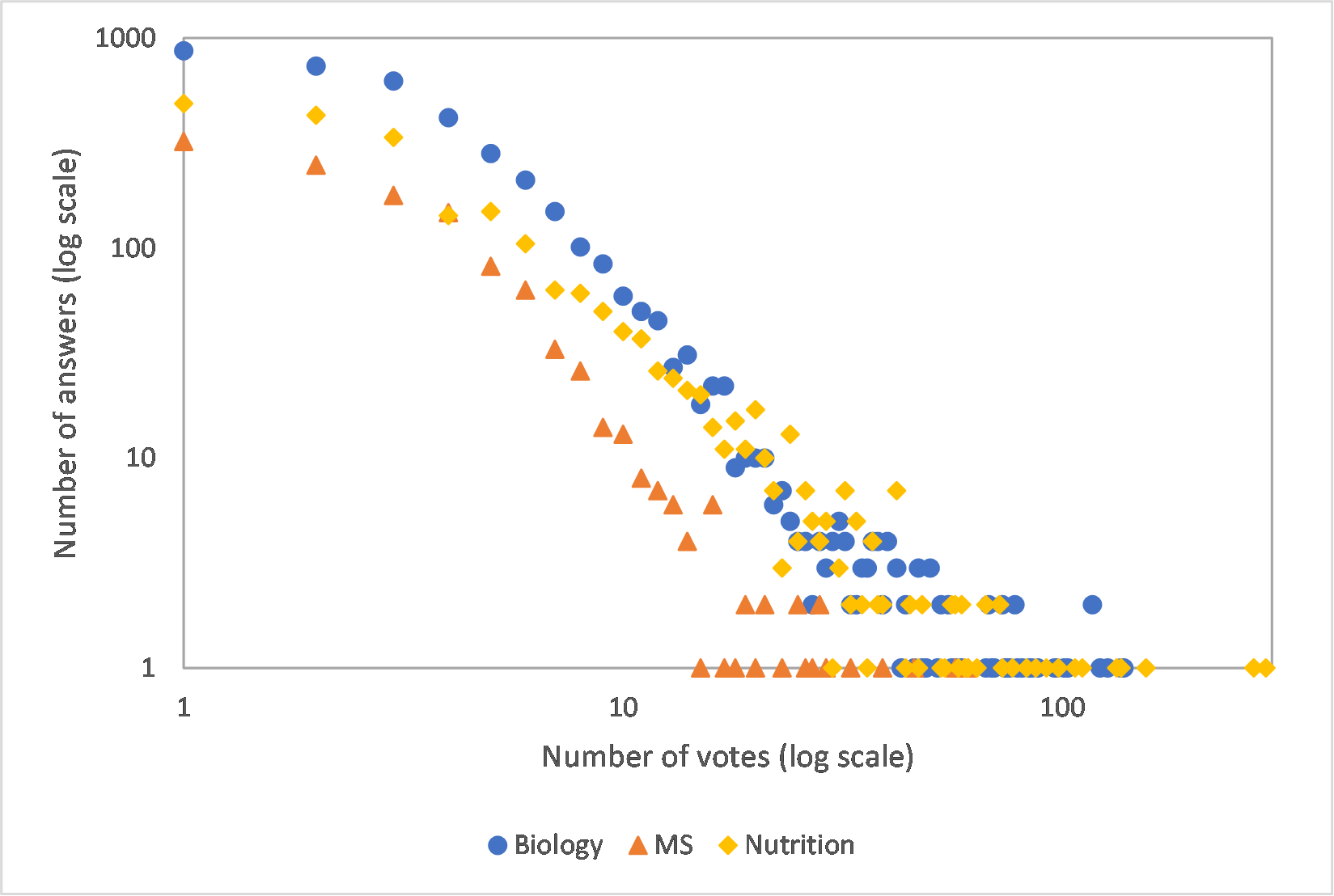}
\caption{Votes distribution of the three datasets.} \label{fig:distvote}
\end{figure*}

\begin{figure*}
\centering
\includegraphics[width=0.8\linewidth]{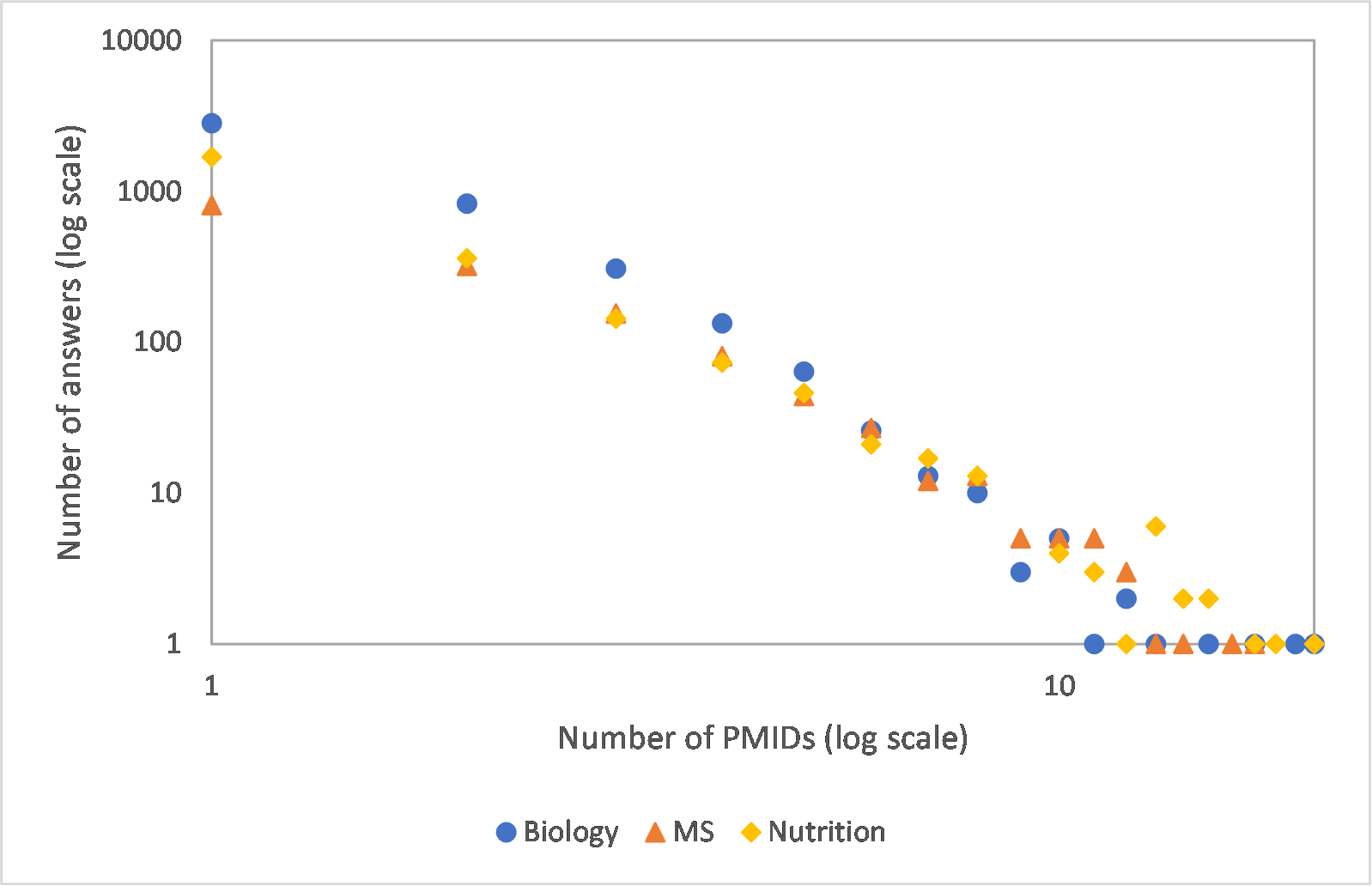}
\caption{PMID count distribution of the three datasets.} \label{fig:distpmid}
\end{figure*}

Each answer has a number of votes associated with it, as well as a list of PMIDs, retrieved at the timestamp shown in Table \ref{table:corpus}.
Fig. \ref{fig:distvote} shows the distribution of each dataset in terms of the number of votes on the answers, while Fig. \ref{fig:distpmid} shows the distribution of the number of PMIDs per answer.
We can see a similar distribution of votes as the one reported by \cite{Dalip2013}.
The MS.SE subset is the smallest of the three forums; hence it also does not have answers with as many votes as the other two.
However, it has a considerable number of answers with many PMIDs, as can be seen in Fig. \ref{fig:distpmid}.
The r/nutrition subset has a higher average number of PMIDs per answer than Biology.SE.
However, many answers had links to other sources of information, such as blog posts and Wikipedia pages.
We also observed that these communities did not always value answers with a large number of citations because more succinct answers are easier to understand.

\subsection{Q-article similarity}
To confirm if the articles provided as answers are actually relevant to the questions, we calculated the cosine similarity between each question title and respective article title.
We used the vectors from the \verb|en_vectors_web_lg| package of Spacy, which are based on a vocabulary of 1M words and 300d vector representations.
These vectors were calculated using the GloVe algorithm~\cite{pennington2014glove} on the Common Crawl dataset~\footnote{\url{https://commoncrawl.org/the-data/}}.
Then we calculated the similarity of each question text with an answer article from a random question of the same dataset.
We calculated the mean and standard deviation of the Q-article pairs and Q-random article pairs, as well as the t-test statistic, for each dataset (Table \ref{table:sim}).
The p-value for every dataset was <0.001, showing that the articles provided by the crowd are significantly more similar to the question than random articles.

\begin{table}[ht]
\centering
\caption{Mean and standard deviation of the similarity between question and respective answer articles, compared with a random answer.}
\label{table:sim}
\begin{tabular}{lllllll}
\hline
                   & \multicolumn{2}{c}{Q-A} & \multicolumn{2}{c}{Q-random A} & \\ %\hline
                   & Mean & $\sigma$  & Mean & $\sigma$    & p-value          \\ \hline
Biology.SE         & 0.7191     & 0.1045 & 0.6287     & 0.1156 & \textless{}0.001 \\ %\hline
MS.SE              & 0.7334     & 0.0993 & 0.6551     & 0.1100  &  \textless{}0.001  \\ %\hline
r/nutrition        & 0.7172     & 0.1029 & 0.6820     & 0.1100  & \textless{}0.001 \\ \hline
\end{tabular}
\end{table}

\section{Search engine evaluation}

As a baseline evaluation of the dataset, we attempted to retrieve the same PubMed documents that were referenced by the answers, using various search engines.
This way, we can observe if a search engine could obtain the same documents as those provided by the users.

Since we do not have a score for each PMID but only for the answer as a whole (which may contain several PMIDs), we have no way to rank the PMIDs from the same answer.
Nevertheless, the number of votes can provide a measure of the relevance of all PMIDs.
On the other hand, it is also difficult to evaluate questions for which we only have a small number of PMIDs, regardless of the number of votes.
For this reason, we experimented with limiting both the minimum number of votes (Table~\ref{table:votes}) and the number of PMIDs associated with the answers (Table~\ref{table:pmids}) while performing a baseline evaluation. %, and its effect on the MAP score.

As explained in Section \ref{sec:comparison}, we retrieved the top 100 articles given by the query likelihood with Dirichlet term smoothing, which is the default scorer of Galago. 
Overall, we can see that increasing the minimum number of PMIDs can lead to higher MAP scores, although this results in a smaller corpus.
The maximum MAP score is achieved when selecting only questions with at least 10 PMIDs for Biology.SE, 5 PMIDs for MS.SE, and 4 PMIDs for r/nutrition.
We were also able to obtain higher MAP scores by filtering by the number of votes, on the MS.SE and r/nutrition datasets.
Ideally, a balance should be found between the minimum number of votes and PMIDs and the size of the corpus.

% update
\begin{table}[ht]
\centering
\caption{Effect of the minimum number of votes per answer on the MAP score and corpus size. The highest score of each subset is bolded.}\label{table:votes}
%\resizebox{\columnwidth}{!}{%
\begin{tabular}{lllllll}
\hline
&   \multicolumn{2}{c}{\textbf{Biology}}  &      \multicolumn{2}{c}{\textbf{MS}} &  \multicolumn{2}{c}{\textbf{Nutrition}} \\ \hline
Min \# Votes     & \# Qs & MAP & \# Qs     & MAP  & \# Qs & MAP \\ \hline %\Xhline{2\arrayrulewidth}
1  & 3714 & 0.0240          & 1332 & 0.0347          & 1963 & 0.0132          \\
2  & 2896 & \textbf{0.0260} & 1114 & 0.0343          & 1550 & 0.0126          \\
3  & 2202 & 0.0251          & 813  & 0.0377          & 1173 & 0.0136          \\
4  & 1614 & 0.0240          & 577  & 0.0311          & 882  & 0.0145          \\
5  & 1211 & 0.0259          & 408  & 0.0324          & 748  & 0.0147          \\
6  & 943  & 0.0239          & 270  & 0.0269          & 616  & 0.0141          \\
7  & 736  & 0.0222          & 191  & 0.0268          & 517  & 0.0143          \\
8  & 593  & 0.0213          & 130  & 0.0331          & 459  & 0.0147          \\
9  & 496  & 0.0214          & 98   & 0.0401          & 402  & \textbf{0.0158} \\
10 & 415  & 0.0226          & 73   & \textbf{0.0403} & 354  & 0.0151         \\ \hline
\end{tabular}
%}
\end{table}

% updated
\begin{table}[ht]
\centering
\caption{Effect of the minimum number of PMIDs per question on the MAP score and corpus size. The highest score of each subset is bolded.}\label{table:pmids}
\resizebox{\columnwidth}{!}{%
\begin{tabular}{lllllll}
\hline
 &   \multicolumn{2}{c}{\textbf{Biology}}  &      \multicolumn{2}{c}{\textbf{MS}} &  \multicolumn{2}{c}{\textbf{Nutrition}} \\ \hline
Min \# PMIDs & \# Qs & MAP & \# Qs     & MAP  & \# Qs & MAP \\ \hline %\Xhline{2\arrayrulewidth}
1  & 3961 & 0.0237          & 1383 & 0.0359          & 2109 & 0.0133          \\
2  & 1508 & 0.0313          & 688  & 0.0441          & 764  & 0.0144          \\
3  & 661  & 0.0355          & 383  & 0.0453          & 395  & 0.0187          \\
4  & 335  & 0.0448          & 216  & 0.0509          & 228  & \textbf{0.0190} \\
5  & 163  & 0.0603          & 127  & \textbf{0.0550} & 153  & 0.0083          \\
6  & 88   & 0.0829          & 78   & 0.0459          & 108  & 0.0075          \\
7  & 51   & 0.1200          & 48   & 0.0504          & 77   & 0.0098          \\
8  & 36   & 0.1462          & 34   & 0.0359          & 55   & 0.0135          \\
9  & 21   & 0.1731          & 23   & 0.0299          & 43   & 0.0153          \\
10 & 18   & \textbf{0.2019} & 19   & 0.0287          & 36   & 0.0171         \\ \hline
\end{tabular}
}
\end{table}

Finally, we studied two other variables on the baseline evaluation: the document retrieval method and the text used as a search query (title, body, or answer).
Table~\ref{table:comparison} shows the results of this study, using each subset of the BiQA corpus.
PubMed search obtained lower MAP scores, even though their method is an improvement over the BM25 scoring method.
Their search engine is tuned for user queries, so one possible explanation is that questions easily answered by a PubMed search are less likely to be posted on Q\&A forums.
Furthermore, in every case, using the body text results in lower scores, and the best scores are achieved using the answer text.
This is expected as the answer text introduces concepts that were missing from the question text and body.

\begin{table}[h!]
\centering
\caption{Effect of search method and query text on the MAP scores, using the BiQA corpus.}\label{table:comparison}
%\resizebox{\columnwidth}{!}{%
\begin{tabular}{lllll}
\hline
                           & \textbf{Section}    & \textbf{Biology.SE} & \textbf{MS.SE} & \textbf{r/nutrition} \\ \hline %\Xhline{2\arrayrulewidth}
\multirow{3}{*}{Pubmed}    & Title  &    0.0066 & 0.0103 & 0.0040        \\ \cline{2-5} 
                           & Body       & 0.0052 & 0.0103 &  0.0031      \\ \cline{2-5} 
                           & Answer     & 0.0787 & 0.0859 &   0.0641     \\ \hline
\multirow{3}{*}{BM25}      & Title      & 0.0267 & 0.0358 & 0.0139 \\ \cline{2-5} 
                           & Body       & 0.0207 & 0.0298 & 0.0098       \\ \cline{2-5} 
                           & Answer & 0.1410 & 0.1571 & 0.0476       \\ \hline
\multirow{3}{*}{Dirichlet} & Title      & 0.0237 & 0.0359 & 0.0133       \\ \cline{2-5} 
                           & Body       & 0.0179 & 0.0285 & 0.0087       \\ \cline{2-5} 
                           & Answer & 0.1370 & 0.1744 & 0.0597       \\ \hline
\end{tabular}
%}
\end{table}

\section{Evaluation on the BioASQ6b dataset}

We applied the whole BiQA corpus to a deep learning QA system to study its impact on this type of system, merging the questions of the three subsets into a single corpus, without any restrictions of number of votes or PMIDs.
We explored two scenarios: using BiQA in conjunction with the BioASQ6b training set and using just the BiQA corpus to train the model.
The results of both experiments, as well as the results we obtained using only the BioASQ6b training data, are shown in Table~\ref{table:bioasq}, for each test batch.

\begin{table*}[h!]
\centering
\caption{MAP scores obtained with the AUEB system using only the BioASQ6b training set, using both BioASQ6b and BiQA, and using only the BiQA corpus to train the model.}\label{table:bioasq}
%\resizebox{0.7\linewidth}{!}{%
\begin{tabular}{cccccc}
\hline
Test batch & BioASQ only & +BiQA &  $\Delta$       & BiQA only &    $\Delta$     \\  \hline %\Xhline{2\arrayrulewidth}
1 & 0.2221      & 0.2235 & -0.0014 & 0.2125  & 0.0096 \\ \hline
2 & 0.2267      & 0.2231 & 0.0035  & 0.2025  & 0.0241 \\ \hline
3 & 0.2415      & 0.2436 & -0.0021 & 0.2279  & 0.0136 \\ \hline
4 & 0.1686      & 0.1712 & -0.0026 & 0.1680  & 0.0006 \\ \hline
5 & 0.1340      & 0.1355 & -0.0015 & 0.1254  & 0.0086 \\ \hline
\end{tabular}
%}
\end{table*}

We obtained higher MAP scores by adding the BiQA corpus to the BioASQ6b training set on 4 out of 5 batches, although these were only marginal improvements.
We can see that our approach obtains MAP scores close to the ones obtained by the AUEB system using only the BioASQ training data.
In fact, in 3 out of 5 batches, the difference is less than 0.01. 
These results show that the BiQA corpus could effectively be used as training data for biomedical QA, and expert curation could lead to better results.
%%%%%%%%%%%%%%%%%%%%%%%%%%%%%%%%%%%%%%

\section{Discussion}

Comparing Tables \ref{table:votes} and \ref{table:pmids}, we can see that, for the Biology.SE dataset, the number of PMIDs per answer is a more efficient way to obtain higher MAP scores, since for similar corpus sizes, the MAP score is higher using a PMID threshold.
Even for higher threshold values, which results in small corpus sizes, the Biology.SE dataset obtains higher MAP scores. 
We did not show results for higher values given the small number of documents that remained.
We also did not include answers with less than one vote since there were only 444 answers in this situation and did not change the results of this analysis.
We expected MS.SE and r/nutrition to have the same behavior as Biology.SE on Table~\ref{table:pmids}, in terms of how the MAP score increases with the minimum number of PMIDs.
Since this was not the case, those two datasets proved to be more challenging and may require better query processing and natural language understanding in order to retrieve the correct documents.
For example, multi-label question classification could also improve the scores of this type of task~\cite{wasim2019}. 

We compare the MAP scores obtained on the BiQA corpus with the ones obtained on the BioASQ6 challenge because it has a subtask that consists in document retrieval for QA (Task B Phase A).
BioASQ is organized yearly, and every year there are new batches of questions.
The participating teams are evaluated on each batch separately, obtaining different scores on each one.
While the best score achieved on the 2019 edition was 0.2898 on test batch 3, on test batch 5, the best score was 0.1218.
We also obtained a range of MAP scores on our corpus, in most cases lower than the BioASQ test batches.
Furthermore, our corpus is closer to a real-world scenario since we are using user-submitted questions, which requires natural language understanding to correctly answer them.
One possibility is that users had access to a search engine before posting a question and did not find an answer, therefore increasing the complexity of the questions.
Novel systems tuned for this type of data can obtain better scores, especially considering that more data is made available each day as users post questions and answers on those forums. 

We were able to train a deep learning model with our corpus that obtained MAP scores similar to a model trained on a corpus annotated by experts. 
One possible explanation for these results is that our corpus has more questions, and these questions were retrieved from various sources.
Furthermore, adding the questions from our corpus to the model trained on the BioASQ6b training data led to slightly better MAP scores on four test batches.
These results highlight the importance of having more datasets for biomedical QA in order to improve the existing systems.

Our approach has some limitations when compared to expert annotations.
Since Q\&A forums are anonymous, there is no accountability of the answers, and personal biases may be more apparent than within a group of expert annotators.
This approach is also dependent on how engaged the users are within the community: for example, r/nutrition has more users and a higher maximum number of votes than the other two, but it has fewer answers that we could link to PubMed articles, resulting in a lower average number of PMIDs.
Another limitation is that there is no context for each citation used to answer.
In some cases, an article is used as an example and not necessarily to answer the question.
Our framework treats every article associated with an answer the same way because we do not know the rank of each article within an answer, so it cannot distinguish which ones are more relevant.

We select the biomedical domain given its complexity and because it is not usually the focus of QA systems.
Although some biomedical QA systems have been proposed \cite{galko2018biomedical}, these are limited in comparison to general ones.
Biomedical-specific approaches are essential due to the relevance of this topic to the society.
On the web, there are frequent health-related questions that are not easily answered but could be clarified through an efficient retrieval of the relevant literature.
Questions from the BiQA corpus that were answered with several citations were related to vaccines, genetically modified food products, and meat consumption, for example, topics for which much information can be found in the scientific literature.

Our framework could potentially be applied to other communities where biomedical questions are made, for example, Quora, or social networks such as Twitter and Facebook.
We selected PubMed as the reference library since it is widely used by the biomedical community and can be easily accessed.
	 could be applied to obtain the answer span of the selected documents~\cite{qiu2020,park2019}.

\subsection{Comparison to other corpora}

\begin{table*}[h!]
\centering
\caption{Comparison of BiQA to other Biomedical and User QA corpora. DR - Document Retrieval; QA - Question Answering.}\label{table:corporacomparison}
\begin{tabular}{llllll}
\hline
Corpus   & Domain          & Task              & Size & Method                & Ref               \\ \hline
BiQA     & Biomedical      & DR/QA             & 7k   & Information retrieval &       -            \\ \hline
BioASQ   & Biomedical      & DR/QA             & 3k   & Expert annotation     &       ~\cite{tsatsaronis2015overview}            \\ \hline
MEDIQA   & Clinical        & QA                & 208  & Expert annotation     & \cite{mediqa2019-overview}    \\ \hline
emrQA    & Clinical        & QA                & 455k & Template              & \cite{pampari-2018-emrqa}        \\ \hline
PubMedQA & Biomedical      & QA                & 211k & Template              & \cite{jin2019pubmedqa}      \\ \hline
Cong2008 & General         & Text classication & 3M   & Information retrieval & \cite{Cong2008} \\ \hline
Shah2010 & General         & Ranking           & 120  & Crowdsourcing         & \cite{Shah:2010:EPA:1835449.1835518}  \\ \hline
AmazonQA & Product reviews & QA                & 923k & Information retrieval & \cite{gupta2019amazonqa} \\ \hline
\end{tabular}
\end{table*}

Table~\ref{table:corporacomparison} provides a summary of existing resources related to User QA and Biomedical QA that can be compared with our approach. 
In comparison to other existing corpora for Biomedical QA, ours is the first to automatically obtain user questions and documents associated with them.
Corpora used in community challenges such as BioASQ  and MEDIQA requires questions and answers obtained from user experts, while PubMedQA and emrQA use synthetically generated questions.
These methods, although useful for some purposes such as training models, lack the diversity that user-generated questions will provide, while expert datasets are limited in size.
Our approach overcomes these two limitations.
In comparison to other User QA corpora, our approach focused on the biomedical domain, and as such, fewer questions can be extracted from the web~\cite{Cong2008, gupta2019amazonqa}.
However, our method takes advantage of the user-interaction aspect of web forum, therefore not requiring crowdsourcing techniques such as the one used by \cite{Shah:2010:EPA:1835449.1835518}.

\section{Conclusion}

This manuscript presents an effective open-source framework to generate corpora for QA document retrieval using Q\&A forums, demonstrated in the form of the BiQA corpus.
These forums are a useful resource for QA systems since the questions are real-world examples of user questions, and the answers are crowdsourced from multiple users.
To the best of our knowledge, this is the first attempt at generating this type of dataset for QA.
The task of document retrieval is a crucial step of a QA system, so these systems must optimize this step according to the user needs.
We demonstrated the feasibility and performance of the framework and the BiQA corpus, namely their relevance for training and evaluating QA systems, particularly in the biomedical domain, for which resources are more scarce.
The BiQA corpus is composed of 7,453 questions and 14,239 question-article pairs obtained from three different Q\&A forums.
The results obtained with our baseline methods indicate that higher MAP scores can be obtained on questions with a higher number of PMIDs, and more advanced methods as well as manual curation could help retrieve articles for all questions correctly.
We release the dataset and the pipeline developed to extract it from the web to promote other works that tackle the challenges of QA systems that address real-world biomedical questions.
As future work, we will use semantic representations to find similar questions, using word embeddings specific for community QA~\cite{zhou-etal-2015-learning,10.1016/j.knosys.2015.11.002}.
These embeddings could be calculated using the corpora retrieved using our method.

\section*{Acknowledgment}

This work was supported by FCT through project DeST, ref.\ PTDC/CCI-BIO/28685/2017, PhD Scholarship, ref.\ SFRH/BD/145221/2019, and the LASIGE Research Unit, ref.\ UIDB/00408/2020 and ref.\ UIDP/00408/2020.

\bibliographystyle{unsrt}
\bibliography{arxiv}

\end{document}